\documentstyle[prl,aps,twocolumn,epsf]{revtex}

\begin{document}
\draft

\twocolumn[\hsize\textwidth\columnwidth\hsize\csname @twocolumnfalse\endcsname

\title
{\bf Surface diffusion coefficients by thermodynamic integration: Cu on
Cu(100)}

\author{Ghyslain Boisvert,$^1$\cite{byline1} Normand
Mousseau,$^2$\cite{byline2} and Laurent J. Lewis$^1$\cite{byline3}}

\address{$^{1)}$
D{\'e}partement de Physique et Groupe de Recherche en Physique et Technologie
des Couches Minces (GCM), Universit{\'e} de Montr{\'e}al, Case Postale 6128,
Succursale Centre-Ville, Montr{\'e}al, Qu{\'e}bec, Canada H3C 3J7
}

\address{$^{2)}$
Department of Physics and Astronomy, Condensed Matter and Surface Science
Program, Ohio University, Athens, Ohio, 45701, USA
}

\date{\today}

\maketitle

 \begin{center}
 \hspace{11pt}
 \end{center}

\begin{abstract}
The rate of diffusion of a Cu adatom on the Cu(100) surface is calculated
using thermodynamic integration within the transition state theory. The
results are found to be in excellent agreement with the essentially exact
values from molecular-dynamics simulations. The activation energy {\it and}
related entropy are shown to be effectively independent of temperature, thus
establishing the validity of the Arrhenius law over a wide range of
temperatures. Our study demonstrates the equivalence of diffusion rates
calculated using thermodynamic integration within the transition state theory
and direct molecular-dynamics simulations.
\end{abstract}


\vskip2pc
]

\narrowtext

Precise knowledge of diffusion processes is essential to understanding
non-equilibrium phenomena such as nucleation and growth\cite{general}. On
surfaces, for instance, the rates at which particles diffuse determine the
equilibrium shape of islands and, on macroscopic timescales, the morphology
of films. Yet, very little is known of the fundamentals of diffusion.
Diffusion constants, for one, are notably difficult to measure and accurate
data is available only for the simplest mechanisms on a small number of
simple surfaces\cite{surfdiff}. Because diffusion is an activated (Arrhenius)
process (at low enough temperatures \cite{tapio}), small errors in the energy
barriers translate into large uncertainties in the diffusion coefficients,
and thus surface structure. In addition, in order to determine the
pre-exponential factor, several measurements are needed in a range of
temperatures over which the Arrhenius behaviour is expected to hold, which is
not always feasible: in practice, the value of the prefactor is often
prescribed. This is a dangerous state of affairs since diffusion obeys the
Meyer-Neldel compensation law\cite{bly95} --- for a family of related
processes, the prefactor increases exponentially with the activation
barrier\cite{mn}.

On the theory side, the situation is just as difficult. It is necessary, in
order to describe diffusion accurately, to have a proper model for the
interatomic potentials. Semi-empirical models, such as the embedded-atom
method (EAM)\cite{fbd86}, while simple and sometimes remarkably accurate,
lack the transferability and predictive power of first-principles methods.
The latter, however, are subject to size and other limitations, and
uncertainties are difficult to estimate. For instance, even for such a simple
case as diffusion by jumps of Cu adatoms on the Cu(100) surface, experiment
and {\em ab initio} calculations disagree\cite{bl97}; the origin of the
discrepancy remains unclear.

Because of various limitations, the technique used for computing diffusion
rates also is important. The simplest option consists in simulating diffusion
explicitely using molecular dynamics (MD). One advantage of this method is
that {\it a priori} knowledge of the diffusion mechanism is not required.
Such calculations are however too demanding for {\em ab initio} methods.
Also, the simulations have to be carried out at relatively high temperatures
where diffusion is ``active'' on MD timescales; at high temperature, however,
diffusion often proceeds by the combination of several mechanisms, making it
difficult to extract individual contributions. Finally, because of possible
anharmonic contributions, the calculated Arrhenius law may not extrapolate to
low temperatures.

Another option consists in computing directly the activation barrier and the
prefactor using the transition-state theory (TST) and various approximations
\cite{bs97,mb93,kkr97,vd84}. Here, however, the reaction path must be known;
while this might be a limitation for bulk diffusion, it is usually not a
serious problem for surfaces where diffusion is relatively well
characterized. In the context of TST, and given a model for the interatomic
potentials, free-energy calculations, in particular thermodynamic integration
(TI), offer the most accurate route to the study of diffusion processes. In
this approach, the diffusion path is followed step by step, and the free
energy calculated using finite-temperature MD. The procedure works best at
low temperature; at high temperature, indeed, diffusion events are more
frequent and the atoms must be constrained to their equilibrium positions
(see below). In this case, it might be more advantageous to use the explicit
MD approach.

Because the two methods are so different, and cover different temperature
ranges, and because diffusion is an important, difficult, and yet unresolved
problem in most cases, it is of utmost interest to ascertain that they lead
to equivalent results. This question has been addressed previously using
Monte Carlo simulations with restricted dynamics on Lennard-Jones
metals\cite{vd84,krwf96}, but the results were not conclusive: the energy
barriers were found to differ by as much as 35\% and the prefactors by a
factor of $\sim$1.8. Here we reexamine the problem in the case of Cu
diffusion on Cu(100), for which detailed MD simulations with EAM potentials
have recently been reported\cite{bl97}; EAM provides a rather accurate
description of the energetics of Cu. The TI is performed in full using MD,
solving directly the TST equations. We find the explicit MD and the TST/TI
calculations to be in {\em very close agreement} for both the prefactor and
the energy barrier. The free-energy barrier, in addition, is found to depend
linearly on temperature, confirming the validity of the Arrhenius law over a
wide range of temperatures. Our results establish unambiguously the
equivalence of the two methods, thus providing a useful framework for the
calculation of diffusion constants.

In the TST, the rate of reaction from one equilibrium site to another, via a
saddle point, is given by \cite{gle41}
   \begin{equation}
   k=\kappa \cdot k_{\rm TST}, \;\;\;\; k_{\rm TST}=\nu e^{-\Delta W/k_BT},
   \label{kk}
   \end{equation}
where $\kappa$ is the transmission coefficient (or ``recrossing rate'') and
$k_{\rm TST}$ is the TST rate constant. $\Delta W$ is the activation free
energy; the prefactor $\nu$, the frequency at which the reaction is
attempted, is given by
   \begin{equation}
   \nu=\left[\frac{k_BT}{2\pi m}\right]^{1/2}
       \left[\int_{\rm well} e^{-[W(x)-W(x_m)]/k_BT} dx\right]^{-1}.
   \label{nu}
   \end{equation}
The integral in Eq.\ (\ref{nu}) runs between two transition sites a distance
$L$ apart, say from $x_b-L$ to $x_b$, via the equilibrium site at $x_m$.
$W(x)$ is the ``potential of mean force'':
   \begin{equation}
   W(x)=\int_{x_m}^{x} <f(\lambda)>_{\lambda=x'} dx',
   \label{w}
   \end{equation}
where $<f(\lambda)>$ is the mean force that must be applied in order to
constrain the particle at position $\lambda$ along the reaction path;
evidently $<f>$ is zero if $x=x_m$ or $x=x_b$. $W$ can be obtained
numerically by calculating the mean force at several points along the
diffusion path using constrained MD\cite{CicRyc}.

\begin{center}
\begin{table}
\caption{
Comparison between TI and MD results for the jump (J) and exchange (X)
diffusion activation barriers $\Delta E$ (in eV) and rate prefactors
$\Gamma_0$ (in THz); also given are the entropy $\Delta S$ (in $k_B$) and
the
static energy barrier, $\Delta E(0)$. Estimated errors are given in
parenthesis.
}
\label{fits}
\begin{tabular}{lcccccc}
          & $\Delta S$ & $\Delta E$ & $\Delta E$ & $\Delta E(0)$ &
$\ln\Gamma_0$ & $\ln\Gamma_0$ \\ \tableline
          &  (TI)      &    (TI)    &    (MD)    &               &
(TI)     &      (MD)     \\ \tableline
J  & 1.1(0.2) & 0.51(0.02) & 0.49(0.01) & 0.50 & 2.9(0.2) & 3.0(0.2) \\
X  & 4.9(0.6) & 0.74(0.02) & 0.70(0.04) & 0.73 & 6.5(0.6) & 6.1(0.7) \\
\end{tabular}
\end{table}
\end{center}

\vspace{-0.7cm}
\begin{figure}
\vspace*{-0.25in}
\epsfxsize=3.2in \epsfbox{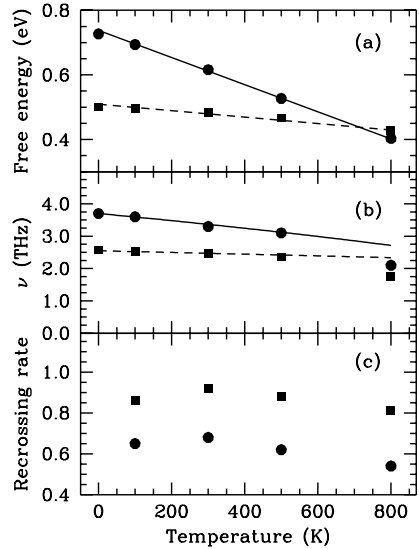}
\vspace{0in}
\caption{
(a) Activation free energy vs temperature for jumps (squares, dashed line)
and exchanges (circles, full line); the lines are linear fits to the
finite-temperature points. (b) Attempt-to-diffuse frequencies vs
temperature;
the lines are the predictions of the simple model discussed in the text.
(c)
Transmission coefficients vs temperature.
\label{fig1}
}
\end{figure}

The TI calculations were carried out using MD and 
EAM potentials. As in Ref.\ \cite{bl97}, the  surface  was modeled 
by a slab consisting of eight layers,
each containing 64 atoms, with the bottom two fixed in their equilibrium
lattice positions; periodic boundary conditions were applied in the two
lateral directions. We investigate here the four temperatures 100, 300, 500,
and 800 K; this will permit a comparison with our earlier MD calculations,
which covered the range 650--850 K\cite{bl97}. Most calculations were done in
the $NVT$ ensemble, using a Nos{\'e} thermostat to control the
temperature\cite{nose}; however, we have also done some calculations in the
$NVE$ ensemble to assess the effect of the thermostat. At each point along
the reaction path, the system was first equilibrated for 48 ps, then
statistics accumulated for a further 120 ps. At the highest temperatures, the
atoms lying close to that undergoing diffusion were attached to their
equilibrium positions with harmonic springs. Several values of the spring
constant were examined and the mean force obtained by extrapolating to
zero\cite{pc92}.

The transmission coefficient $\kappa$ is given by
   \begin{equation}
   \kappa=<\Theta[x(+t)-x_b]-\Theta[x(-t)-x_b]>_{t\gg\tau_{\rm vib}}
   \label{kappa}
   \end{equation}
where $\tau_{\rm vib}$ is a time characteristic of atomic vibrations and
$\Theta$ is the Heaviside step function. $\kappa$ was obtained by averaging
over 100 different initial configurations, taken at 1.2 ps intervals from a
MD run with the adatom constrained at the saddle point. Each of these was run
for 1.2 ps both backward and forward in time\cite{roux91}.

We plot in Fig.\ \ref{fig1}(a) the activation free energies $\Delta W$ as a
function of temperature for both mechanisms possible 

\begin{figure}
\vspace*{-0.5in}
\epsfxsize=3.2in \epsfbox{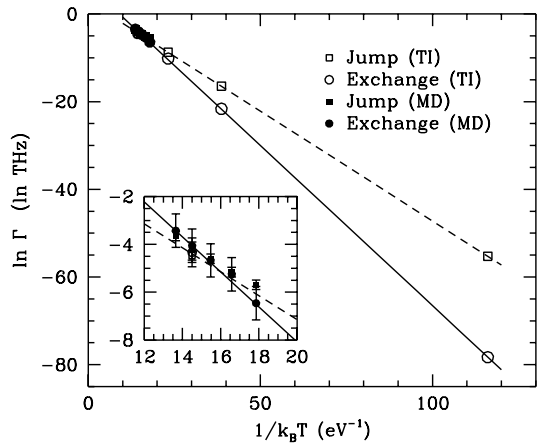}
\vspace{0in}
\caption{
Diffusion rates vs inverse temperature; the lines are fits to an Arrhenius
law; estimated errors are smaller than the size of the symbols. Inset:
Comparison between MD and TST/TI results in the temperature range relevant
to
the MD data.
\label{fig2}
}
\end{figure}

\noindent on this surface, viz.,
jump and exchange; the static (0 K) values are also indicated. In both cases,
$\Delta W$ is very well represented by a linear function of temperature,
i.e., $\Delta W = \Delta E - T \Delta S$, where $\Delta E$ and $\Delta S$ are
both, {\em effectively}, temperature-independent. The values of $\Delta E$
and $\Delta S$ are listed in Table \ref{fits} along with the corresponding
values for $\Delta E$ from the direct MD simulations. For both diffusion
processes, {\em the TI and MD barriers are in excellent agreement with the
static barriers}.

We note from Fig.\ \ref{fig1}(a) that, in spite of the fact that it has a
larger activation barrier, exchange diffusion is more favourable than jump
diffusion above 800 K or so. This is a manifestation of the Meyer-Neldel rule
\cite{bly95,mn}; the prefactor for exchanges is much larger (20 times ---
cf.\ Table \ref{fits}) than that for jumps, thus compensating for the smaller
activation term. Compensation is so efficient that the process with a larger
barrier becomes dominant at sufficiently high temperature. Thus, even at low
temperature, where anharmonic effects are small, multi-phononic contributions
to the entropy cannot be ignored.

The attempt-to-diffuse frequencies for the two processes are displayed in
Fig.\ \ref{fig1}(b). They depend only very weakly on temperature. The
deviations at 800 K might reflect some error in the TI calculations at high
temperatures. However, even taking such possible errors into account, the
observed temperature dependence is insignificant relative to the exponential
activation term and thus, for all practical purposes, the prefactor for
diffusion via a {\em given} mechanism can be taken as constant.

The (slow) variation of $\nu$ with temperature can be understood in terms of
the following simple anharmonic model: Taking $W(x)$ to be of the form
$\alpha x^2/2 - \beta x^3/3$, with $W(x_m) \equiv 0$ and $\Delta W =
W(x_b)-W(x_m)$ (so that $\alpha = 6\Delta W/x_b^2$ and $\beta=6\Delta
W/x_b^3$) one easily finds from Eq.\ (\ref{nu}) (neglecting the anharmonic
term in the evaluation of the integral) that, in the low temperature limit,
$\nu(T) = \nu_0 \sqrt{\Delta E - T \Delta S}/\sqrt{\Delta E}$ with $\nu_0 =
\nu(0)$. We see from Fig.\ \ref{fig1}(b) that the TI data is very well fitted
by this model up to about 500 K; as expected, this approximation is no longer
valid at higher temperature. Note that the slight temperature dependence will
be negligible, again, on an Arrhenius plot. The differences in the
$\nu$-values for the two processes arise, to a large extent, from
``geometrical'' differences. For the above model we also have $2\pi\nu =
\sqrt{6 \Delta W/mx_b^2}$, where $m$ is the mass of the diffusing entity ---
$m_{\rm Cu}$ for jump and $m_{\rm Cu}/2$ for exchange (motion of a dimer with
respect to its center of mass); taking $x_{b,X} = 1.6$ \AA\ for an exchange
(roughly $a/2$, with $a=3.61$ \AA\ the lattice parameter) and $x_{b,J} = 1.3$
\AA\ for a jump [$a/(2\sqrt{2})$], we find, indeed, $\Delta W_J/x_{b,J}^2
\approx \Delta W_X/x_{b,X}^2$.

The transmission coefficient is the probability that a diffusion event
actually takes place once the saddle point is reached. For both mechanisms,
$\kappa$ depends relatively little on temperature, as can be seen in Fig.\
\ref{fig1}(c). For jumps, the transmission coefficient is about 0.9, and thus
has little effect on the diffusion prefactor. For exchanges, $\kappa$ is
close to 0.6, and the effect is slightly more important.

The diffusion rate is the product of transition rate, transmission
coefficient and number of equivalent reaction paths. (The diffusion {\em
constant} is obtained from the diffusion {\em rate} by multiplying by a
geometrical factor.) For both jumps and exchanges, there are four equivalent
paths and we thus have [cf.\ Eq.\ (\ref{kk})]:
   \begin{equation}
   \Gamma=4\kappa\nu e^{-\Delta W/k_BT} \equiv \Gamma_0 e^{-\Delta E/k_BT},
       \Gamma_0=4\kappa\nu e^{\Delta S/k_B}
   \label{gamma}
   \end{equation}
The TI results for $\Gamma$ are presented in Fig.\ \ref{fig2}. The data are
extremely well fitted by an Arrhenius law at all temperatures, even as large
as 800 K. The resulting values of $\Delta E$ are nearly identical to those
determined earlier by fitting to the free energies. The slight temperature
dependence of the attempt-to-diffuse frequencies and the transmission
coefficients has, as anticipated, no visible effect on the Arrhenius
barriers. The values of the prefactors $\Gamma_0$, which we return to below,
are listed in Table \ref{fits}.

There has been some concerns that the thermostat in $NVT$ simulations might
lead to sizable errors in free-energy calculations (see, e.g., Ref.\
\cite{kkr98}). In order to test this, we have carried out some TI
calculations for both jump and exchange at 500 K, using both $NVT$ and $NVE$
algorithms. Differences were found to be insignificant --- at most 0.007 eV
on free energies and 0.01 THz on diffusion rates --- well within numerical
uncertainties.

In the inset of Fig.\ \ref{fig2}, finally, we compare closely the TI results
with the MD simulations. The former covers the range 0--800 K, while the
latter is for 650--850 K. {\em The TI and MD calculations are found to be in
complete agreement for both diffusion mechanisms over the whole temperature
range}. This establishes without ambiguity that the two different
computational schemes complement one another exactly. In addition, our
calculations demonstrate that the range of validity of the Arrhenius law can
extend over a much wider range of temperatures than is normally assumed.

Free-energy calculations of the barriers for jump diffusion on Cu and Ag
(100) surfaces based on the {\em harmonic} approximation to TST have been
reported recently\cite{kkr97}. The calculations were carried out using the
same model potentials (EAM) as in the present study; yet, for Cu jump on
Cu(100), a prefactor 10 times smaller than that found here was obtained.
Numerical error cannot be totally excluded as the cause for this discrepancy,
but the consistency between our TI and MD results strongly suggests that this
is not the case. Rather, it is more likely a problem with methodology: the
harmonic and quasi-harmonic approximations neglect the multiphononic
contributions which affect deeply the thermodynamic functions, especially
prefactors, giving rise, as we have seen earlier, to such effects as the
Meyer-Neldel law\cite{bml98}.

It has been claimed by many authors (see for example Refs.\ \cite{kkr97},
\cite{hsjn93} and \cite{sanders92}) that the entropy $\Delta S$ and the
energy barrier $\Delta E$ depend on temperature. Our results provide no
evidence for this. The separation of the different terms in Eq. (\ref{kk}) is
somewhat arbitrary and largely a matter of definition. The simplest
expression for $\Gamma$, viz.\ $\Gamma=\Gamma_0 \exp(-\Delta E/k_BT)$, where
$\Gamma_0$ (and thus $\Delta S$) as well as $\Delta E$ are {\em effectively
independent of temperature}, is able to account very precisely for both the
TI and the MD data over the full range of temperatures considered. Indeed,
the entropy term, after dividing by $k_BT$, merely renormalizes the prefactor
[cf.\ Eq.\ (\ref{gamma})].

We have reported a detailed comparison of the rates for jump and exchange
self-diffusion on Cu(100) as obtained from full thermodynamic integration and
direct molecular-dynamics simulations. We find the two methods to be in {\em
perfect agreement} over a wide range of temperatures. Our results clearly
demonstrate that a simple representation of the diffusion rate in terms of a
static energy barrier (which defines the activation term) and a
temperature-independent entropy (which defines the prefactor), as they appear
in the usual transition state theory, accounts fully for the dynamics of
isolated adatoms. Furthermore, the present study clearly demonstrates the
equivalence of the diffusion constants obtained within TST/TI and from direct
MD simulations.

{\it Aknowledgements ---}
We are grateful to Beno{\^\i}t Roux for useful advice in the initial stages
of this project. This work is supported by grants from the Natural Sciences
and Engineering Research Council (NSERC) of Canada and the ``Fonds pour la
formation de chercheurs et l'aide {\`a} la recherche'' (FCAR) of the Province
of Qu{\'e}bec. One of us (G.B.) is thankful to NSERC and FCAR for financial
support.


\begin{references}

\bibitem[\star]{byline1} e-mail address: boisver@physcn.umontreal.ca

\bibitem[\dag]{byline2} e-mail address: mousseau@helios.phy.ohiou.edu

\bibitem[\ddag]{byline3} e-mail address: lewis@physcn.umontreal.ca

\bibitem{general} E. Kaxiras, Comp. Mat. Sci. {\bf 6}, 158 (1996).

\bibitem{surfdiff} G. L. Kellogg, Surf. Sci. Rep. {\bf 21}, 1 (1994).

\bibitem{tapio} T. Ala-Nissila and S.C. Ying, Prog. Surf. Sci. {\bf 39}, 227
(1992).

\bibitem{bly95} G. Boisvert, L.J. Lewis, and A. Yelon, Phys. Rev. Lett. {\bf
75}, 469 (1995).

\bibitem{mn} W. Meyer and H. Neldel, Z. Tech. Phys. {\bf 12}, 588 (1937).

\bibitem{fbd86} S.M. Foiles, M.I. Baskes, and M.S. Daw, Phys. Rev. B {\bf
33}, 7983 (1986).

\bibitem{bl97} G. Boisvert and L.J. Lewis, Phys. Rev. B {\bf 56}, 7643
(1997).

\bibitem{bs97} M. Bockstedte and M. Scheffler, Z. Phys. Chemie {\bf 200},
195-207 (1997).

\bibitem{mb93} D. Maroudas and R. A. Brown, Phys. Rev. B {\bf 47}, 15562
(1993).

\bibitem{kkr97} U. K{\"u}rpick, A. Kara, and T.S. Rahman, Phys. Rev. Lett. {\bf
78}, 1086 (1997).

\bibitem{vd84} A. F. Voter and J. D. Doll, J. Chem. Phys. {\bf 80}, 5832
(1984).

\bibitem{krwf96} P.V. Kumar, J. S. Raut, S.J. Warakomski, and K.A. Fichthorn,
J. Chem. Phys. {\bf 105}, 686 (1996).

\bibitem{gle41} S. Glasstone, K. J. Laidler, and H. Eyring, {\it Theory of
Rate Processes}, McGraw-Hill, New York, 1941.

\bibitem{CicRyc} G. Ciccotti and J. P. Ryckaert, Comp. Phys. Rep. {\bf 4},
345 (1986).

\bibitem{nose} S. Nos\'e, Mol. Phys. {\bf 52}, 255 (1984).

\bibitem{pc92} E. Paci and G. Ciccotti, J. Phys.: Condens. Matter {\bf 4},
2173 (1992).

\bibitem{roux91} B. Roux and M. Karplus, J. Phys. Chem. {\bf 95}, 4856
(1991).

\bibitem{kkr98} U. K{\"u}rpick, A. Kara, and T.S. Rahman, Phys. Rev. Lett. {\bf
80}, 204 (1998).

\bibitem{bml98} G. Boisvert, N. Mousseau, and L.J. Lewis, Phys. Rev. Lett.
{\bf 80}, 203 (1998).

\bibitem{hsjn93} L.B. Hansen, P. Stoltze, K.W. Jacobsen, and J.K. N{\o}rskov,
Surf. Sci. {\bf 289}, 68 (1993).

\bibitem{sanders92} D. E. Sanders {\it et al.} Surf. Sci. {\bf 264}, L169
(1992).

\end{references}
\end{document}